\documentstyle[prl,twocolumn,epsf,floats,aps]{revtex}
\begin{document}
\draft

\twocolumn[\hsize\textwidth\columnwidth\hsize\csname @twocolumnfalse\endcsname

\title{The two-dimensional bond-diluted quantum Heisenberg model at the\\
classical percolation threshold}

\author{Anders W. Sandvik}
\address{Department of Physics, University of Illinois at Urbana-Champaign,
1110 West Green Street, Urbana, Illinois 61801}
\date{September 15, 1999}

\maketitle

\begin{abstract}
The two-dimensional antiferromagnetic $S=1/2$ Heisenberg model with 
random bond dilution is studied using quantum Monte Carlo simulation 
at the percolation threshold ($50\%$ of the bonds removed). Finite-size 
scaling of the staggered structure factor averaged over the largest connected
clusters of sites on $L \times L$ lattices shows that long-range order 
exists within the percolating fractal clusters in the thermodynamic limit. 
This implies that the order-disorder transition driven by bond-dilution 
occurs exactly at the percolation threshold and that the exponents are 
classical. This result should apply also to the site-diluted system.
\end{abstract}


\vskip2mm]

During the past decade, questions related to the destruction
upon doping of the antiferromagnetic order in the high-$T_c$ cuprate materials
have motivated extensive studies of quantum critical phenomena in 
two-dimensional (2D) antiferromagnets \cite{sachdev}. The 2D Heisenberg 
model on a square lattice can be driven through an order-disorder transition
\cite{chn} by, e.g., introducing frustrating interactions 
\cite{frust} or by dimerizing the lattice \cite{dimer}. It has also been 
believed that a non-trivial phase transition could be achieved by diluting 
the system, i.e., by randomly removing either sites 
\cite{wan,behre,manousakis,yasuda,castro} or bonds (nearest-neighbor
interactions) \cite{awsmv}. The site-dilution problem is of direct relevance 
to the cuprates doped with nonmagnetic impurities, such as Zn or Mn 
substituted for Cu \cite{doping}. Early numerical work indicated that the 
long-range order vanishes in the Heisenberg model with nearest-neighbor 
interactions when a fraction $p^* \approx 0.35$ of the sites are 
removed  \cite{behre}. Various analytical treatments have given results 
for $p^*$ ranging from $0.07$ \cite{yasuda} to $0.30$ \cite{castro}. 
These hole concentrations are below the classical percolation threshold 
$p_{\rm cl} \approx 0.407$ \cite{percdens}, and hence the phase transition 
would be caused by quantum fluctuations. However, in a recent paper Kato 
{\it et al.} reported quantum Monte Carlo simulations of larger lattices
at lower temperatures than in previous work and concluded that the critical 
site-dilution is exactly the percolation density; $p^*=p_{\rm cl}$ 
\cite{kato}. Based on their simulations, they also argued that the critical 
behavior nevertheless is not classical, but that the fractal clusters at 
$p_{\rm cl}$ are quantum critical with algebraically decaying correlation 
functions. This leads to non-classical critical exponents. Most surprisingly,
the simulations indicated that the exponents depend on the magnitude 
of the spin.

In this Letter we use a quantum Monte Carlo method to study the related
bond-diluted system exactly at the percolation threshold, i.e., with
$50$\% of the bonds randomly removed. In order to determine the nature of
the ground state at this point --- quantum critical, classically critical, or
quantum disordered --- we study the magnetic properties of the largest 
clusters of connected sites on $L \times L$ lattices with $L$ up to 
$18$. We find clear evidence that these clusters, which in the thermodynamic
limit are fractal with fractal dimension $d=91/48$ \cite{stauffer}, are 
antiferromagnetically ordered. This implies that the order-disorder transition
driven by bond-dilution occurs exactly at the percolation threshold and that
the critical exponents are classical. 

We have chosen to study the bond dilution problem because it is 
numerically more tractable than site 
dilution --- since the bond percolation threshold $p_{\rm cl}=1/2$ it
can be realized exactly on the finite lattices we work with. However, the 
fractal dimension and the critical exponents are the same for classical bond 
and site percolation \cite{marro}, and our conclusions should therefore hold
true also for the site-diluted system. We argue that the reason for the 
disagreement with the previous results by Kato {\it et al.} \cite{kato} is 
that they did not use sufficiently low temperatures in their simulations
--- extremely low temperatures are required for reaching the ground state 
even for the relatively small system sizes we use here.

We consider the $S=1/2$ Heisenberg Hamiltonian on a square lattice 
with $N=L \times L$ sites;
\begin{equation}
H = \sum_{b=1}^{2N} J(b) {\bf S}_{i(b)} \cdot {\bf S}_{j(b)}.
\end{equation}
The bonds $b$ connect nearest neighbor sites $i(b),j(b)$ with interaction 
strength $J(b)=J$ for $N$ randomly selected bonds and $J(b)=0$ for the 
remaining $N$ bonds. We use an efficient finite-temperature quantum Monte 
Carlo method based on the ``stochastic series expansion'' approach 
\cite{sse1,sse2} to study systems with $L=4,6,\ldots, 18$. In order to 
reach the ground state, we successively increase the inverse temperature
$\beta \equiv J/T$ until all quantities of interest have converged. We find 
that $\beta$ as high as $\approx 10^4$ is required for the largest lattice 
we have considered. Another important concern when studying random systems 
is the equilibration of the simulations. One would like to average over as 
many random configurations as possible within given computer resources. 
Ideally, one would then perform only short simulations for each realization
(typically, even a quite short simulation of a given configuration results 
in a statistical error smaller than the fluctuations between different 
configurations). However, there is a minimum length of a simulation set by 
the time needed to equilibrate it, and when carrying out short simulations 
it is important to have some way to verify that the correct equilibrium 
distribution indeed has been reached. We use the following scheme to check 
for both equilibration and temperature effects: For each bond configuration
we carry out simulations at inverse temperatures $\beta = 2^nL$, 
$n=0,1,\ldots,n_{\rm max}$. Starting with $n=0$, we perform two runs 
for every $\beta$, each with $N_{\rm e}$ updating steps for equilibration
and $N_{\rm m}=2N_{\rm e}$ steps for measuring physical quantities (for the 
definition of a ``step'', see Ref.~\cite{sse2}). The second run is a direct 
continuation of the first one, so that the effective number of equilibration
steps is four times that for the first run. An agreement between the results
of these two runs is then a good indication that the simulation has 
equilibrated. For the subsequently lower temperatures (increasing $n$) we 
always start from the last Monte Carlo state generated at the previous 
temperature. The convergence of the simulations using the $\beta$-doubling 
procedure will be illustrated with som results below. 

In the thermodynamic limit, the system at $p_{\rm cl}$ will be spanned by 
infinite clusters of fractal dimension $d=91/48$ \cite{stauffer}. The 
existence of a nontrivial (quantum) critical point is determined by the 
magnetic properties of these clusters. If they are long-range ordered, the 
critical point of the order-disorder transition driven by bond dilution will 
be exactly at $p_{\rm cl}=1/2$ as in the classical case, and the critical 
exponents will be the classical ones. If the fractal clusters
are critical, i.e., their spin-spin correlations decay with a power-law 
as suggested by Kato {\it et al.} \cite{kato}, the critical point is still 
at $p_{\rm cl}$ but the exponents are different and non-trivial. A quantum 
critical point with $p^* < p_{\rm cl}$ would imply an exponential decay of 
the spin-spin correlations within the fractal clusters at $p_{\rm cl}$. 
In order to determine which of these scenarios apply, we have calculated the 
magnetization squared of the largest connected cluster of sites for a large 
number of random bond configurations on lattices of different linear size $L$. 
As $L \to \infty$ this procedure gives the ordered moment of the fractal 
clusters of interest. Denoting the largest cluster by $C$, the 
$z$-component of the staggered magnetization squared is given by
\begin{equation}
m_C^2 = \left\langle \left ( 
{1\over N_C} \sum_{i \in C} (-1)^{x_i+y_i} S^z_i \right )^2
\right\rangle,
\label{mc2}
\end{equation}
where $N_{C}$ is the number of spins in the cluster and the brackets 
indicate both the quantum mechanical expectation value and an average over 
bond configurations. We also consider the full staggered structure factor
\begin{equation}
S_\pi = \left\langle {1\over N} \left ( 
\sum_{i=1}^N (-1)^{x_i+y_i} S^z_i \right )^2
\right\rangle ,
\label{spi}
\end{equation}
which involves all the spins of the lattice and was used by Kato {\it et al.}
\cite{kato} to study the critical behavior of the site-diluted system. The 
number of random bond configurations used in the averages presented here
ranges from $\approx 10^3$ for $L=18$ to $\approx 10^4$ for $L=4$.

\begin{figure}
\centering
\epsfxsize=8.2cm
\leavevmode
\epsffile{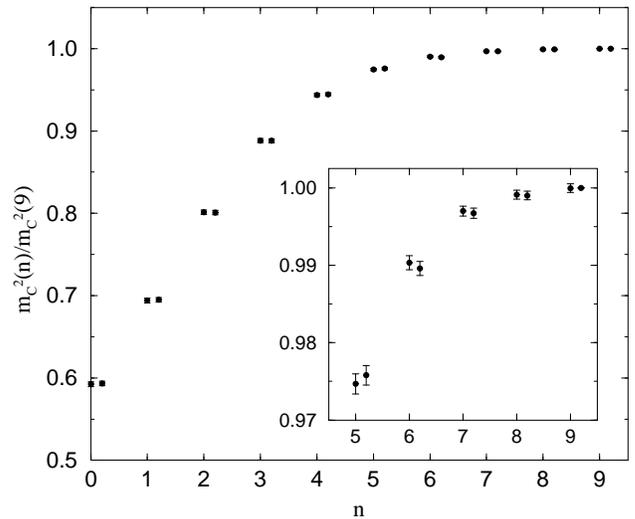}
\vskip1mm
\caption{Staggered magnetization squared for the largest cluster of connected
spins on lattices with $L=18$, averaged over $\approx 1000$ bond configurations
and normalized by the result obtained in the second simulation at the lowest
temperature. The index $n$ corresponds to inverse temperature according 
to $\beta=2^nL$. The points for the second of the two equal-$\beta$ runs 
have been shifted to the right in order to enable a better comparison between 
the data sets. The inset shows the low-temperature data on a more detailed 
scale.}
\label{fig1}
\end{figure}

In Fig.~\ref{fig1} we show results illustrating the equilibration scheme.
The disorder-averaged $m_C^2$ is graphed versus the index $n$ (specifying
the inverse temperature $\beta=2^nL$ as described above) for $L=18$. In 
order to reduce effects of fluctuations among bond realizations and more
clearly show the relative effects of equilibration times and temperature,
we have normalized the data to the result of the second run at the highest
$\beta$ used ($n_{\rm max}=9$, corresponding to $\beta=9216$) and estimated 
the statistical errors using the bootstrap method \cite{bootstrap}. The number 
of equilibration and measurement steps were $N_{\rm e} =500$ and 
$N_{\rm m}=1000$. Within error bars, there are no differences between any of
the equal-$\beta$ runs, and we therefore conclude that the simulations are well
equilibrated. As an additional check, for $L=16$ and smaller we have 
also carried out simulations using $N_{\rm e} =250$ and $N_{\rm e}=1000$ 
(keeping $N_{\rm m}=2N_{\rm e}$). For $N_{\rm e}=250$ we do see small but 
statistically significant differences between the equal-$\beta$ runs, but 
the results of the second run are always consistent with data obtained 
using the longer equilibrations. Hence, we believe that our results are
free of detectable effects of insufficient equilibration. All the
results to be presented below were averaged using the second of 
the equal-$\beta$ runs only. 

\begin{figure}
\centering
\epsfxsize=8cm
\leavevmode
\epsffile{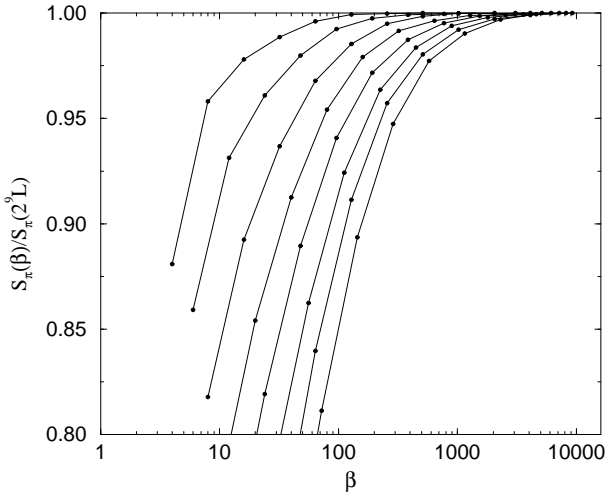}
\vskip1mm
\caption{Disorder-averaged staggered structure factor vs inverse
temperature for systems of linear size $L=4,6,\ldots, 18$ (from top).
The data has been normalized by the results obtained at the highest
$\beta$ for each system size ($\beta=2^9L$).}
\label{fig2}
\end{figure}

Fig.~\ref{fig1} shows that very low temperatures are needed to
converge to the ground state even for a lattice of relatively modest size.
Since there are no statistically significant differences between the $L=18$ 
results at $\beta=4608$ and $9216$, and the asymptotic approach to the ground
state should be exponential, $\beta=9216$ should give the ground state to
within error bars. In Fig.~\ref{fig2} we show normalized results for $S_\pi$ 
as a function of $\beta$ for all the lattices we have studied. It is clear 
that using a fixed $\beta=500$ for all system sizes, as was done in 
Ref.~\cite{kato}, leads to a significant systematic error (assuming that 
site-diluted systems are similarly affected by temperature, which can be 
expected up to some factor of order $1$). Fig.~\ref{fig2} shows that the 
relative deviation from the ground state grows rapidly with $L$ at fixed 
$\beta$ and is $\approx 3$\% for $L=18$ at $\beta=500$. The largest lattice 
considered in Ref.~\cite{kato} was $L=48$, for which our results suggest 
that $\beta=500$ could lead to an error of more than $10$\%. The results
to be discussed below are all for $\beta=2^9L$.

The reason for the very high $\beta$-values needed to converge to the ground
state is most likely that localized moments can form in the irregular clusters
of connected spins. These moments interact with each other with a strength
which decreases rapidly with increasing separation, thus leading to closely
spaced energy levels. The typical level spacing should decrease faster with 
increasing $L$ than the $1/L$ behavior suggested in Ref.~\cite{castro}. The 
temperature effects should be the largest exactly at the percolation threshold
and for all hole concentrations they lead to an underestimation of the ordered
moment. Hence, the result that there is a substantial order in the system 
close to the percolation threshold \cite{kato} should remain valid despite 
such effects. 

\begin{figure}
\centering
\epsfxsize=8.2cm
\leavevmode
\epsffile{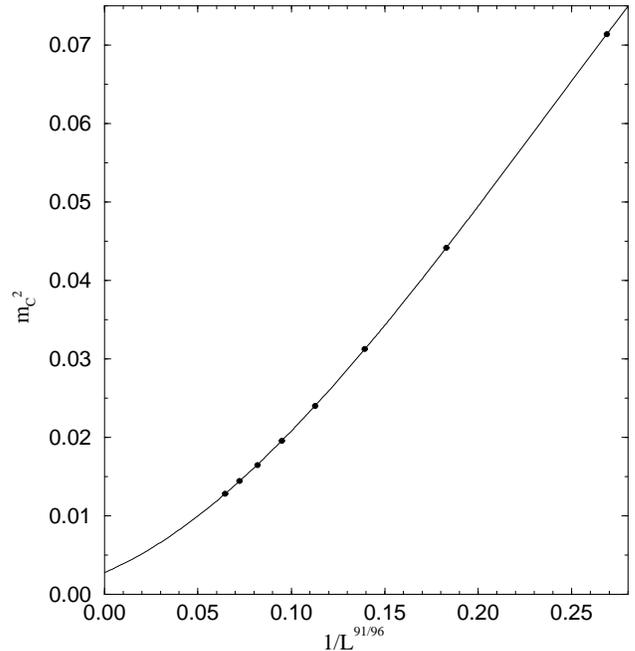}
\vskip1mm
\caption{Finite-size scaling of the disorder-averaged staggered magnetization
squared of the largest connected clusters on $L \times L$ lattices. Statistical
errors are smaller than the symbols. The curve is a fit to a cubic 
polynomial.} 
\label{fig3}
\end{figure}

In order to definitely determine whether indeed $p^*=p_{\rm cl}$
and whether or not the exponents are the classical ones, we here study the 
lattice-size dependence of the cluster order parameter squared; 
Eq.~(\ref{mc2}). As a finite-size scaling ansatz, we use a 
simple generalization of the known scaling law for the sublattice 
magnetization $m$ for the pure 2D Heisenberg model. In that case the leading 
size-correction to $m^2$ is $\sim 1/N^{1/2}$ \cite{huse}, which can be seen 
clearly in numerical data \cite{num2d}. Since the average number of spins in 
the fractal clusters of the diluted system depends asymptotically on 
$L=N^{1/2}$ according to $\langle N_C \rangle \sim L^d$, with $d$ the fractal 
dimension quoted above, we here assume a leading size correction 
$\sim 1/L^{d/2}$. Fig.~\ref{fig3} shows our data for $m_C^2$ graphed 
versus this variable. The results appear to be consistent with the 
ansatz, although in order to fit all the data points we have to use
a polynomial cubic in $1/L^{d/2}$ (a cubic polynomial is needed also to fit 
data for the pure 2D Heisenberg model, but the corrections to the linear 
behavior are much smaller \cite{num2d}). This fit has a $\chi^2$ per degree 
of freedom of $\approx 0.6$ and gives the full staggered magnetization 
$M_d = \sqrt{3m_C^2} \approx 0.09$ for the infinite fractal clusters (the 
factor $3$ accounts for rotational averaging). Hence, the order on the 
$d$-dimensional fractal lattice is as high as $\approx 30$\% of the 2D 
staggered moment \cite{num2d}.

\begin{figure}
\centering
\epsfxsize=8.2cm
\leavevmode
\epsffile{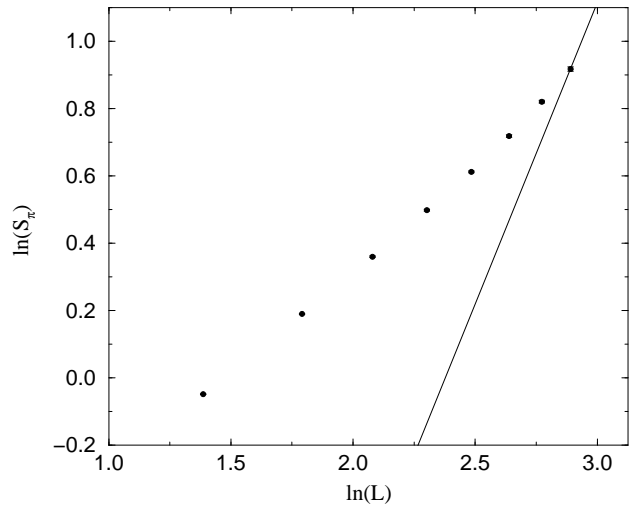}
\vskip1mm
\caption{The logarithm of the staggered structure factor vs the logarithm
of the system size. The line has the slope $43/24$ expected in the asymptotic
classical scaling regime.} 
\label{fig4}
\end{figure}

Finally, we discuss results for the staggered structure factor of the
full lattice; Eq.~(\ref{spi}). In Fig.~\ref{fig4} we graph $\ln{(S_\pi)}$
versus $\ln{(L)}$, which should show an asymptotic linear scaling behavior.
There is a clear upward curvature as the lattice size increases, and it is 
clear that we are still far from the scaling regime. A curvature is also 
seen in the data presented by Kato {\it et al.} for small systems \cite{kato}, 
but for larger sizes a linear behavior was discerned. As we have discussed 
above, the structure factor calculated for large lattices in Ref.~\cite{kato}
was likely substantially under-estimated due to temperature effects, and 
the apparent non-classical scaling
behavior is then an artifact. Since the results in Fig.~\ref{fig3} show that 
there is long-range order in the fractal clusters, the growth of $S_\pi$ 
must asymptotically be given by classical percolation theory, i.e., 
$S_\pi \sim L^{2d-2}$, where $2d-2 = 43/24$ \cite{stauffer}. The 
very large corrections to scaling evident in Fig.~\ref{fig4} can be understood
as resulting from a significant reduction with increasing $L$ of the staggered
structure factor per site of the fractal clusters ($m_C^2$), as seen in 
Fig.~\ref{fig3}. Classically, $m_C^2$ is independent of system size and the 
scaling of $S_\pi$ is therefore determined solely by the increase in size of 
the clusters as $L$ increases. In the quantum case, the size 
dependence of $S_\pi$ is dominated by this geometric effect only for system 
sizes sufficiently large for the relative size-correction to $m_C^2$ to
be small. For our largest lattice, the relative size correction is still
more than a factor four. Considering the extremely low temperatures required
to study the ground state of large lattices, it will be very difficult to 
numerically observe the asymptotic classical scaling regime for $S=1/2$.

In summary, we have presented numerical results showing that the fractal
$S=1/2$ Heisenberg clusters at the classical bond-percolation density have 
long-range antiferromagnetic order. This implies that the order-disorder 
transition driven by bond-dilution occurs exactly at the percolation density
and that the critical exponents are classical. This should hold true for
any spin $S$, and since classical bond and site percolation are equivalent
in terms of the fractal dimension and the critical exponents \cite{marro}, 
our conclusions should apply also to the site-diluted model. The quantum 
mechanical corrections to the asymptotic scaling behavior are very large for
lattice sizes that can be studied with today's computers, making direct 
observation of the critical behavior difficult. We have also discussed 
temperature effects and shown that extremely low temperatures are needed to 
study the ground state, likely due to the presence of weakly interacting 
localized moments.

It would be interesting to study a diluted system including frustration.
If the clean system is already close to a quantum critical point, non-magnetic
impurities may be able to drive it into a quantum disordered phase. The
fact that Zn or Mn doping of the cuprates destroys the long-range order
well before the classical percolation threshold \cite{doping} clearly 
indicates that these materials cannot be described by a randomly diluted 
Heisenberg model with nearest-neighbor interactions only.

This work was supported by the NSF under grant No.~DMR-97-12765. Some of 
the calculations were carried out using the Origin2000 
system at the NCSA.

\null\vskip-12mm\null

\end{document}